# Extension of the Wu-Jing equation of state (EOS) for highly porous materials: calculations to validate and compare the thermoelectron model

Geng Huayun,[*] Wu Qiang, Tan Hua, Cai Lingcang and Jing Fuqian

Laboratory for Shock Wave and Detonation Physics Research, Southwest Institute of Fluid Physics, P. O. Box 919-102, Mianyang Sichuan 621900, People's Republic of China

In order to verify and validate the newly developed thermoelectron equation of state (EOS) model that is based on the Wu-Jing (W-J) EOS, calculations of shock compression behavior have been made on five different porous metals—iron, copper, lead, tungsten, and aluminum—which are commonly used as standards. The model was used to calculate the Hugoniot, shock temperature, sound velocity, and unloading isentrope for these materials and comparisons were made to previous calculations and available data. Based on these comparisons, it is felt that the model provides information in good agreement with the corresponding experimental and theoretical data published previously. This suggests that the new model can satisfactorily describe the properties of shocked porous materials over a wide range of pressure and porosity.

[*] E-mail: genhy@sohu.com

## I. INTRODUCTION

When highly porous materials are subjected to shock compression, a large amount of heat is generated during the compression. If the material is porous enough and the shock intensity is high enough, abnormal Hugoniot behavior is observed because the material expands due to the heating.[1-6] This anomaly manifests itself in the form of a pressure-specific volume (P-V) curve that is multi-valued in volume. This behavior invalidates many EOS models because they do not account for more than one value.[7] This problem can be avoided by using a new method to investigate the thermodynamic variable along isobaric paths.[8-9]

A new EOS that was based on this idea was proposed by Wu and Jing in 1995.[9-10] In a companion paper (Ref.11) we theoretically developed their model further by accounting for the effect of thermoelectrons which are important in the shock compression of highly porous materials. This development led to new forms for the Hugoniot, the shock temperature, and the release isentrope. This paper has been written to compare calculations using the new model with previous modeling and experimental data for highly porous metals.

For reference purposes, the principal equations developed in the companion paper are reproduced here,

$$V_h^{'} = V_h + \frac{R_c}{2-R_c}(V_{00} - V_0) + \frac{\beta T^{'2}}{4P} \quad \text{for the Hugoniot,} \tag{1}$$

$$\frac{dT^{'}}{dP} - \frac{R}{P}T^{'} = \frac{1}{2C_P}\left(V_{00} - V_h^{'} + P\frac{dV_h^{'}}{dP}\right) \quad \text{for the shock temperature,} \tag{2}$$

$$\frac{\partial V_s}{\partial P} = \left(1 - \frac{R}{2}\right)\frac{\partial V_h^{'}}{\partial P} - V_h^{'}\left(\frac{R-2}{2P} + \frac{dR}{RdP}\right) - \frac{R}{2P}V_{00} + V_s\left(\frac{R-1}{P} + \frac{dR}{RdP}\right) \quad \text{for the release isentrope.} \tag{3}$$

Here,

$$\frac{dV_h'}{dP} = \left(1 + \frac{\beta T'^2}{8PV_h}\right)\frac{dV_h}{dP} + \left[2(V_{00} - V_0)/(2 - R_c)^2\right]\frac{dR_c}{dP} + \frac{\beta T'}{2P}\frac{dT'}{dP} - \frac{\beta T'^2}{4P^2}, \tag{4}$$

$$\frac{\partial V_h'}{\partial P} = \left(1 + \frac{\beta T'^2}{8PV_h}\right)\frac{dV_h}{dP} + \left[2(V_{00} - V_0)/(2 - R_c)^2\right]\frac{dR_c}{dP} - \frac{\beta T'^2}{4P^2}, \tag{5}$$

$$P = C_0^2 (V_0 - V_h)/(V_0 - \lambda(V_0 - V_h))^2, \tag{6}$$

$$C_P = C_{P0}\left[1 + (1+Z)^{-2}\right]/2 + 3\beta T'/2, \tag{7}$$

$$R = \frac{R_c R_e}{R_c + R_e} = \frac{R_c}{3R_c + 1}, \tag{8}$$

$$\beta = \beta_0 \left(\frac{V_h}{V_{0K}}\right)^{1/2}. \tag{9}$$

In the above equations, the prime denotes the physical quantity for the porous material and subscripts $h$, $s$, $e$, and $c$ refer to the variables on a Hugoniot, the variables on an isentrope, the variables contributed by electrons, and the variables contributed by crystal lattices, respectively. The symbols $V$, $T$, $P$, $R$, $\beta$, and $C_P$ denote the specific volume, the temperature, the pressure, the W-J EOS parameter, the coefficient of electronic specific heat, and the specific heat at constant pressure of the material, respectively. Using Eqs.(1)~(9), the Hugoniot, the shock temperature, the bulk sound velocity, and the release isentrope can be calculated for high-porosity materials. There are eight independent parameters introduced in this model: $V_{00}$, $V_0$, $V_{0K}$, $\beta_0$, $C_0$, $C_{P0}$, $\lambda$, and $Z$. These refer to the initial specific volume of the porous material, the initial specific volume of the solid matrix at the normal state, the initial specific volume of the solid matrix at 0K, the coefficient of electronic specific heat, the bulk sound velocity, the specific heat at constant pressure of crystals at the normal state, a proportionality factor coming from the linear relationship of the shock and particle velocities, and the solid irrelevance, respectively.[1,5] In addition to these, there are several other parameters which are based on these that must also be determined.

## II. DETERMINATION OF PARAMETERS

In this extended EOS model, an important parameter is the W-J parameter, *R*, which occurs in a number of places in Eqs.(1)~(9). It is important to evaluate this parameter and its derivative with respect to pressure, *dR/dP*, before continuing the discussion. Since $R_e$, the part of *R* for the electrons, is in general a constant, we can put more effort into the determination of $R_c$, the part for the crystals, to evalute *R*. Using an analysis similar to that used in the Mie-Grüneisen EOS development of the vibration model of crystal lattices, $R_c$ can be written as[12]

$$R_c = \frac{d\ln\Theta_D}{d\ln V} \cdot \frac{d\ln V}{d\ln P} = \frac{\gamma P}{K}, \tag{10}$$

where $\Theta_D$ is the Debye temperature and $K$ is the bulk modulus. Eq.(10) can be further developed as

$$R_{DM} = \left[\frac{1}{3} + \frac{V}{2} \cdot \frac{\partial^2 \left(P(V)V^{2/3}\right)/\partial V^2}{\partial \left(P(V)V^{2/3}\right)/\partial V}\right] \cdot \frac{d\ln V}{d\ln P} \tag{11}$$

for the Einstein solid model with one dimensional oscillators, and

$$R_f = \frac{V}{2} \cdot \frac{\partial^2 \left(P(V)V^{4/3}\right)/\partial V^2}{\partial \left(P(V)V^{4/3}\right)/\partial V} \cdot \frac{d\ln V}{d\ln P} \tag{12}$$

for the Free-Volume model.

It is difficult to decide which of these two different expressions should be selected for the calculations. Because of this, in the case of high-porosity materials, we have adopted an arithmetic average of them which appears to work well. That is, the crystal W-J parameter $R_c$ is taken as

$$R_c = \frac{R_{DM} + R_f}{2} \tag{13}$$

and
$$dR_c = \frac{dR_{DM} + dR_f}{2}. \tag{14}$$

It is more efficient to evaluate this parameter on the solid Hugoniot rather than on the compression line at zero-Kelvin. Hence, the function $P(V)$ or $V(P)$ in Eqs.(11) and (12) should be replaced with the solid

Hugoniot Eq.(6).

Another parameter $Z = lR_g T / \mu C_x^2$ is called the solid irrelevance,[1,5] in which $R_g$ is the universal constant for gas, $\mu$ is the mole mass of the studied material, $C_x$ is the mean velocity of elastic waves (that can be replaced by $C_0$ in general), and $l$ is the anharmonic parameter denoting the anharmonic degree of crystal vibrations.[2,11] Here $T$ is the shock temperature and should be replaced by $T'$ for porous cases.

In this way, all the parameters in the principal equations are completely determined and the properties of shock-compressed porous materials (such as Hugoniot, shock temperature, and sound velocity at high pressures) can all be calculated. The material parameters used in following calculations are listed in Table I. The values of all parameters were independently determined from the solid materials, not from fitting the experimental data of the porous materials.

TABLE I. Parameters for solid materials used in the calculations.

| Materials | $\rho_0$[a] (g/cm$^3$) | $\rho_{0K}$[a] (g/cm$^3$) | $\lambda$[a] | $\beta_0$[a] (erg/g·K$^2$) | $C_{P0}$[b] (cal/g) | $l$ (anhar. para.)[a] | $\mu$ (g)[c] | $C_0$[a] (km/s) |
|---|---|---|---|---|---|---|---|---|
| Fe | 7.85 | 7.96 | 1.92 | 193.9 | 0.091 | 8 | 55.8 | 3.574 |
| Cu | 8.93 | 9.05 | 1.51 | 174.4 | 0.0845 | 9 | 63.5 | 3.91 |
| Pb | 11.34 | 11.56 | 1.47 | 104.5 | 0.0304 | 30 | 106.7 | 2.03 |
| W | 19.2 | 19.31 | 1.23 | 83.85 | 0.0321 | 8 | 184 | 4.04 |
| Al | 2.71 | 2.764 | 1.341 | 415.3 | 0.211 | 6 | 26.97 | 5.392 |

[a]Xu and Zhang (see Ref. 13).

[b]Handa (see Ref. 14).

[c]Jing (see Ref. 1).

## III. CALCULATIONS TO VALIDATE AND COMPARE THE THERMOELECTRON MODEL

### A. Shock temperature

Since Eqs.(1), (3)~(5), and (7) all depend on shock temperature, Eq.(2) is critical. In order to calculate this, Eqs.(1), (4), (6)~(9), and (11)~(14) must be substituted into Eq.(2). This yields a complicated first-order differential equation of shock temperature. In general, this differential equation does not have a rigorous analytic solution.

By numerically integrating this equation with respect to pressure, we have used it to calculate temperature for solid aluminum, porous copper and iron with different initial densities. The calculated temperatures for these three kinds of materials are in good agreement with the corresponding experimental data and some theoretical data,[15-19] as shown in Figs.1 and 2. The calculated shock temperatures (solid lines) at two initial densities for porous copper are shown in Fig.1(a) and for porous iron at two initial densities in Fig.1(b). They can be compared with the theoretical data points obtained by Gryaznov using a non-ideal plasma EOS model (Ref.16) which are plotted. In this paper porosity $m=V_{00}/V_0$, the ratio of the porous material initial specific volume to the solid material specific volume. Obviously, the two models are comparable over the temperature range shown. Here, the "non-ideal plasma model" is based on considering the shocked porous material as a mixture of electrons, atoms, and ions of different charges interacting with one another. The free energy of such a system is split into two parts: 1) the ideal-gas contribution of atoms, ions, and electrons and 2) the inter-particle interactions.[20] The calculated shock temperature vs pressure for solid aluminum and solid iron are the lines shown in Fig.2; Fig.2(a) is for solid aluminum compared with the theoretical data given by Al'tshuler *et al.*(Ref.15), and Fig.2(b) is for solid iron compared with the experimental data of Bass (Ref.17) and Tang (Ref.18) and other theoretical results by McQueen (Ref.19).

There is a good match between the data points and the calculated curves suggesting the model of this paper is consistent with previous models and experimental data for both solid and highly porous materials over a wide pressure range.

Moreover, these figures also affirm the expansion of the applicability of this model for porous materials from near-solid initial densities to highly porous materials up to $m$=20 (Fig.1), and these provide increased understanding of the shock properties of materials under high pressure and high temperature conditions.

## B. Shock Hugoniot

Calculating the shock Hugoniot also involves all of the equations that were used in calculating the shock temperature. Substituting them into Eq.(1) and we obtain an explicit Hugoniot relation for this model. Copper, iron, lead, and tungsten, which are commonly used as standards, have been selected as examples to verify this Hugoniot expression for porous materials. Aluminum is also used as an illustration for the solid case.

The calculated Hugoniots for porous tungsten, iron, lead, and copper with different initial densities, as well as solid aluminum, are compared with the corresponding experimental data published previously,[2-4, 16, 20-21] and are shown in Figs.3~7, respectively. Fig.3 gives the calculated Hugoniots (solid lines) for solid and porous aluminum. The solid line ($m$=1) can be compared to the corresponding experimental and theoretical data given by Al'tshuler (Ref.15). The good agreement is another indication that this extended model is applicable to solid as well as porous materials.

The calculated Hugoniots for porous copper (solid lines) from $m$=1.41 to 10 are compared with the corresponding experimental data of Trunin (Refs.2~4) and Bakanova (Ref.21) and the theoretical results of Gryaznov (Ref.16) in Fig.4. Then, in Fig.5, the calculated Hugoniots for porous lead with different initial densities are compared with the corresponding experimental data of Trunin (Ref.3). Fig.6 shows the calculated Hugoniots for porous tungsten with different initial densities compared with the corresponding experimental data of Trunin (Ref.3). Fig.7 shows the calculated Hugoniots for porous iron compared with the corresponding experimental data of Trunin (Refs.2, 3) and the theoretical results of the non-ideal plasma model of Gryaznov (Ref.16).

The good agreement of the calculated Hugoniots with the experimental data shown in these figures indicates the effectiveness of this model in being able to predict the Hugoniots of porous metals over a wide range of pressure and porosity. The calculated Hugoniots for ultra-porous copper with *m*=7.2, 10 and iron with *m*=10, 20, respectively, compare nicely with the theoretical predictions given by Gryaznov[16] using the non-ideal plasma model as shown in Figs.4 and 7. The calculations using the extended model described in this and the companion paper extend to higher pressures and porosities than the previous model because it has accounted for the effect of thermoelectrons of materials.

## C. Release isentrope and sound velocity

To further validate this EOS model from the point of view of sound velocity and the release isentrope, we must use all the equations from Eq.(1) to (14). The sound velocity is obtained from $C^2 = -V_h^{'2}(\partial V_s/\partial P)_h^{-1}$, where $(\partial V_s/\partial P)_h$ can be deduced from Eq.(3) by setting $V_s = V_h^{'}$, that is

$$\left(\frac{\partial V_s}{\partial P}\right)_h = \left(1 - \frac{R}{2}\right)\frac{\partial V_h^{'}}{\partial P} + \frac{R}{2P}\left(V_h^{'} - V_{00}\right). \qquad (15)$$

By making use of these equations, we can calculate the sound velocities for porous materials. However, since there is a lack of experimental data for high-porosity materials, we have compared the calculation only to solid copper and slightly porous iron in Fig.8. The calculated bulk sound velocities (solid lines) for porous iron, with an initial density of 6.91g/cm$^3$, and solid copper compare well with the experimental data of Li (Ref.22), Al'tshuler (Ref.23), and Meyers (Ref.24), respectively.

We have also calculated the unloading isentropes of shocked porous metals using Eq.(3). Fig.9 shows the calculated release isentropes for shocked porous tungsten (solid lines) for two different shock intensities which took the porous tungsten ($m$=2.16) to pressures of 116GPa and 152GPa, respectively. The experimental data of Gudarenko (Ref.25) are also shown for these two impact shocks. Fig.10 shows the calculated release isentropes for shocked porous copper ($m$=2.41) taken to a pressure of 138GPa in the initial shock and then released. This line can be compared to the experimental data of Zhernokletov (Ref.26). The calculations for both porous tungsten and porous copper fit the data quite well. In these experiments the porous material was shocked to the high-pressure state and then released to zero pressure.

The experimental data for the release isentropes were obtained by impedance matching the shock compressed samples to serial materials with lower shock Hugoniots. The shock velocity in these barrier materials was measured and the states have been transformed to the pressure-specific volume plane for comparison purposes in this paper.

## IV. DISCUSSION

Since nothing about phase transitions has been accounted for in this EOS model, the calculated results of this paper are useable only in regions where there are no phase transitions. In general, there are cusps or discontinuities in the shock Hugoniots in the T-P plane in the regions of a phase transition.[1, 27-28] This model does not account for this but it is possible that the latent heat of phase change could be introduced to correct this.

In Fig.1, our calculations are lower than those obtained in the non-ideal plasma model[16] when the shock temperature is below 10000 Kelvin. The reason for this is that the non-ideal plasma model becomes invalid when the shock temperature is lower than the ionizing temperature of the material and results in a higher than is credible temperature, i.e., the discrepancy between the two models is expected.

The electronic Grüneisen parameter (EGP) of transition metals such as iron is still controversial. Due to the atom's unusual electron shell structure, the electronic specific heat of this metal is much larger than that of the normal metals at zero-pressure. However, under high-pressure conditions, they trend to be approximately equal. Russian researchers have suggested the use of the experimental data to determine the value of electronic Grüneisen parameter. Their suggested value of the EGP for iron, when temperature below 50000$K$, is ~1. We have chosen to use a value of 0.5 in this paper, which is a theoretical value evaluated by the Thomas-Fermi model, because the corresponding theoretical electron specific heat of iron based on the Thomas-Fermi model is used here rather than the experimental value. Based on the calculated results shown in Figs.1 and 7, this was a reasonable thing to do.

To further analyze the Hugoniot part of this extended EOS model for porous materials, we can rewrite Eq. (1) as

$$V_h^{'} = V_H{'} + V_e, \qquad (16)$$

where

$$V_e = \frac{\beta}{4P} T^{'2}. \qquad (17)$$

In these expressions, $V_h^{'}$ is the whole specific volume with $V_H{'}$ the crystal part and $V_e$ the thermoelectron part. Taking porous iron with initial porosities of $m$=10 and 20 as examples, we have calculated the relative contributions of the two terms, $V_H{'}$ and $V_e$, for shocked porous iron up to pressures of 100GPa. The results are plotted in Fig.11 as fractions of the whole. This figure shows that the contribution of the second term, *i.e.* the part contributed by thermoelectrons, quickly increases with increasing pressure and ends in a relatively stable level. This level increases with increasing porosity. On the other hand, the crystal part decreases to a relatively stable level, with the level decreasing as the porosity increases.

## V. CONCLUSIONS

Using the extended W-J EOS model, which was developed in the companion paper (Ref.11) for predicting the Hugoniot relationships of porous materials using the corresponding solid Hugoniot as a reference, the calculated Hugoniots for porous tungsten, copper, iron, lead, and aluminum with different porosities have been determined. These have been compared to available experimental data and data calculated using other models. Good agreement has been demonstrated, validating this model as a useful tool for estimating the shock states of highly-porous shocked materials.

In addition, shock temperatures, sound velocities, and unloading isentropes of shocked porous materials have also been evaluated. The calculated temperatures for porous aluminum, copper, and iron with different initial densities are in good agreement with the corresponding experimental and theoretical data published previously. Sound velocity calculations for solid and slightly porous samples are also good. Since there are no experimental data available for highly porous materials, further validations of the sound velocity calculations are not possible. However, calculated unloading isentropes have been compared with the experimental data for porous copper and tungsten and the results are reasonable, suggesting that this model is doing a credible job of determining states achieved in the shock and those attained during the unloading process.

**Acknowledgements**

This study was financially supported by the National Natural Science Foundation of China under Grant No. 19804010 and Science and Technology Foundation of CAEP under Grant No. 980102.

**Figure captions:**

Fig. 1. Calculated shock temperatures (solid lines) for porous copper (a) and porous iron (b) with different initial densities. These are compared with data obtained by Gryaznov using a non-ideal plasma EOS model (Ref.16). Here $m=V_{00}/V_0$.

Fig. 2. (a) Calculated shock temperature (solid line) for solid aluminum compared with the theoretical data of Al'tshuler *et al.* (Ref.15); the two curves are almost identical. (b) calculated temperature for solid iron compared with the experimental data of Bass (Ref.17), Tang (Ref.18), and the theoretical results of McQueen (Ref.19).

Fig. 3. Calculated Hugoniots (solid lines) for aluminum compared with the corresponding experimental and theoretical data obtained by Al'tshuler for solid aluminum (Ref.15) ($m=V_{00}/V_0$).

Fig. 4. Calculated Hugoniots (solid lines) for porous copper with different initial densities compared with the corresponding experimental data of Trunin (Refs.2, 3, and 4) and Bakanova (Ref.21) and the theoretical results of Gryaznov (Ref.16) ($m=V_{00}/V_0$).

Fig. 5. Calculated Hugoniots (solid lines) for porous lead with different initial densities compared with the corresponding experimental data of Trunin (Ref.3) ($m=V_{00}/V_0$).

Fig. 6. Calculated Hugoniots (solid lines) for porous tungsten with different initial densities compared with the corresponding experimental data of Trunin (Ref.3) ($m=V_{00}/V_0$).

Fig. 7. Calculated Hugoniots (solid lines) for porous iron with different initial densities compared with the corresponding experimental data of Trunin (Refs.2 and 3) and the theoretical results of Gryaznov (Ref.16) ($m=V_{00}/V_0$).

Fig. 8. Calculated bulk sound velocities (solid lines) of porous iron and solid copper using Eq.(15). The calculations are compared to the solid copper experimental data of Meyers (Ref.24) and Al'tshuler (Ref.23).

Calculations for porous iron ($\rho_{00}$=6.91g/cm$^3$) are compared to the experimental data of Li (Ref.22).

Fig. 9. Calculated unloading isentropes of shocked, porous tungsten (solid lines) ($m$=2.16) compared with the experimental data of Gudarenko (Ref.25). The two sets of data are for two loading pressures, 116 and 152GPa. The Hugoniot shown has been calculated for porous tungsten ($m$=2.16).

Fig. 10. Calculated unloading isentropes for shocked, porous copper (solid line) compared with the experimental data of Zhernokletov (Ref.26). The porous copper was shocked to a pressure of 138GPa and then released. The Hugoniot shown has been calculated for porous copper ($m$=2.41).

Fig. 11. The relative contributions of the two terms, $V_H'$ and $V_e$, on the right-hand side of Eq.(16) to the whole specific volume of the system for porous iron with $m$=10 and 20. The solid lines are the crystal contribution ($V_H'$) and the dot-dashed lines are the thermoelectron contribution ($V_e$). These curves were generated by calculating the shock states for shocks from 1 to 100GPa; they represent Hugoniot type data for the two contributions.

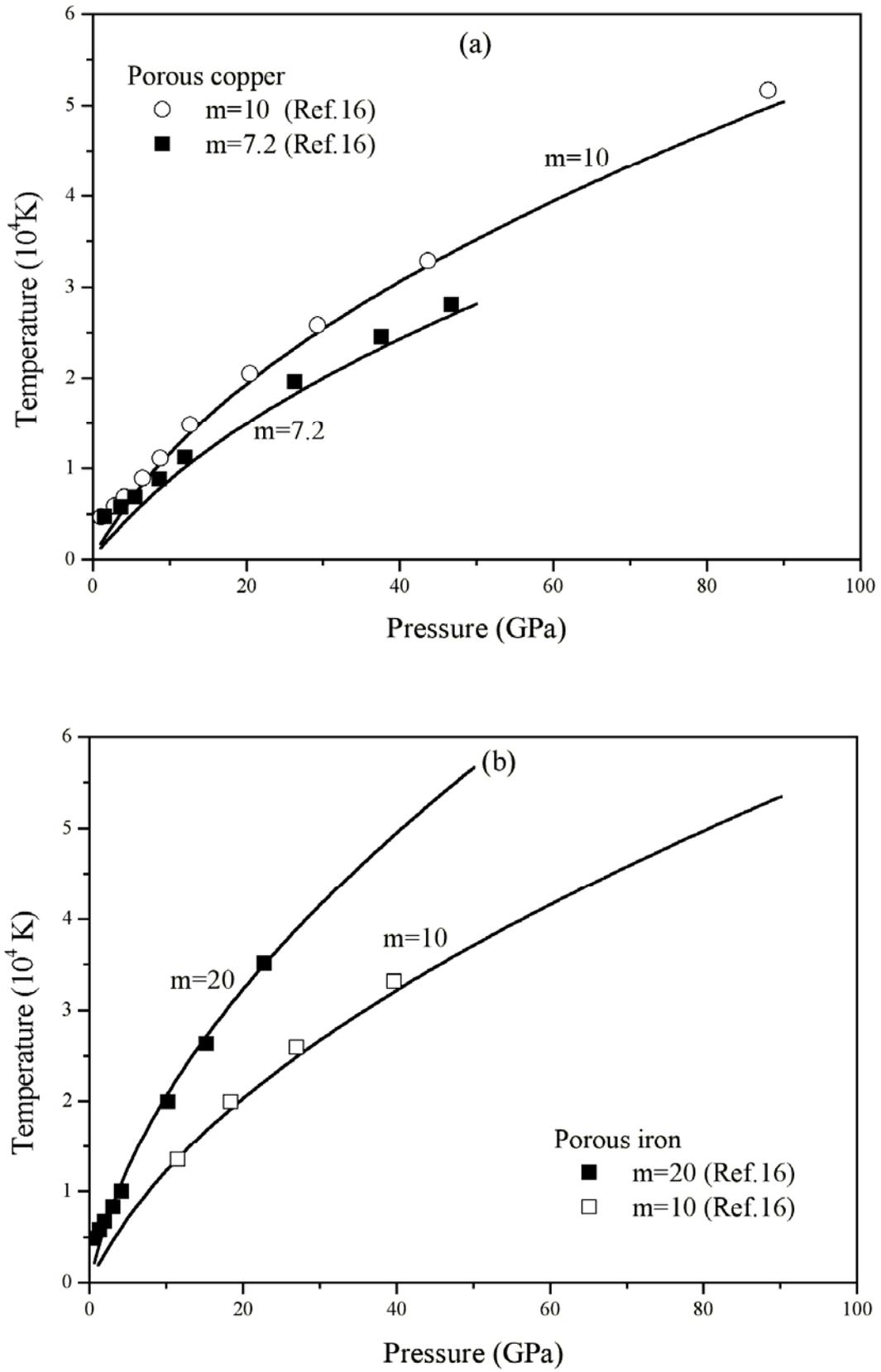

Fig.1, by Geng Huayun *et al.*

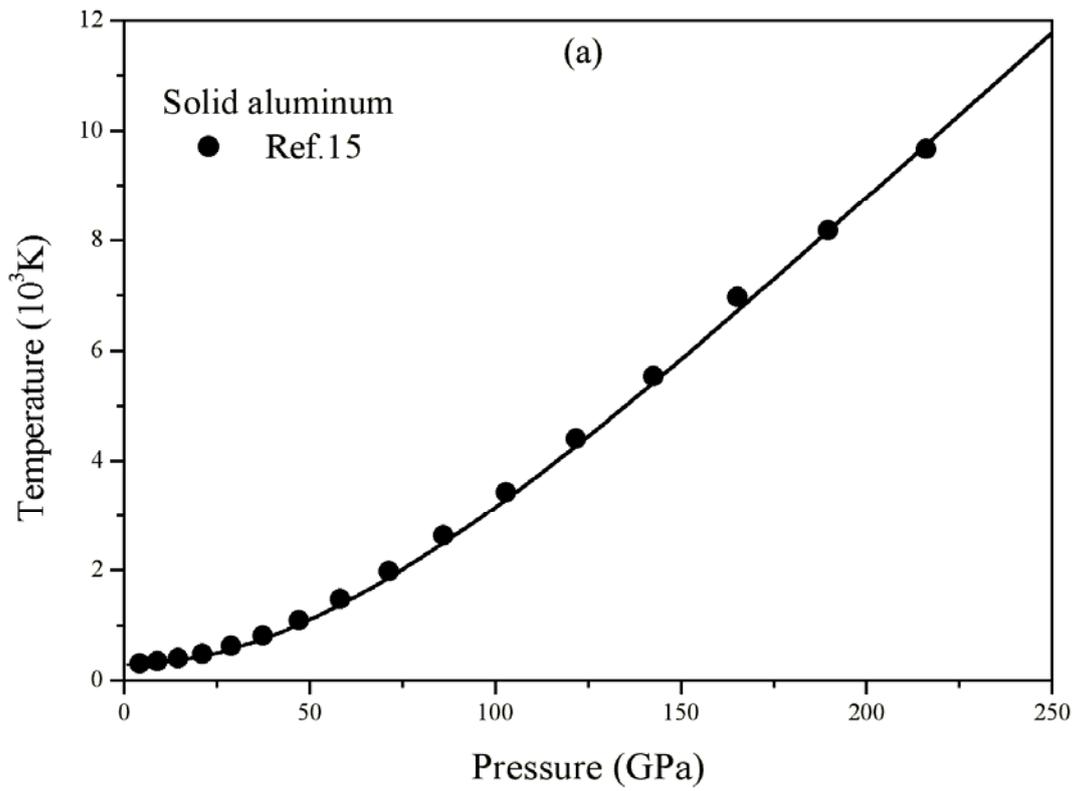

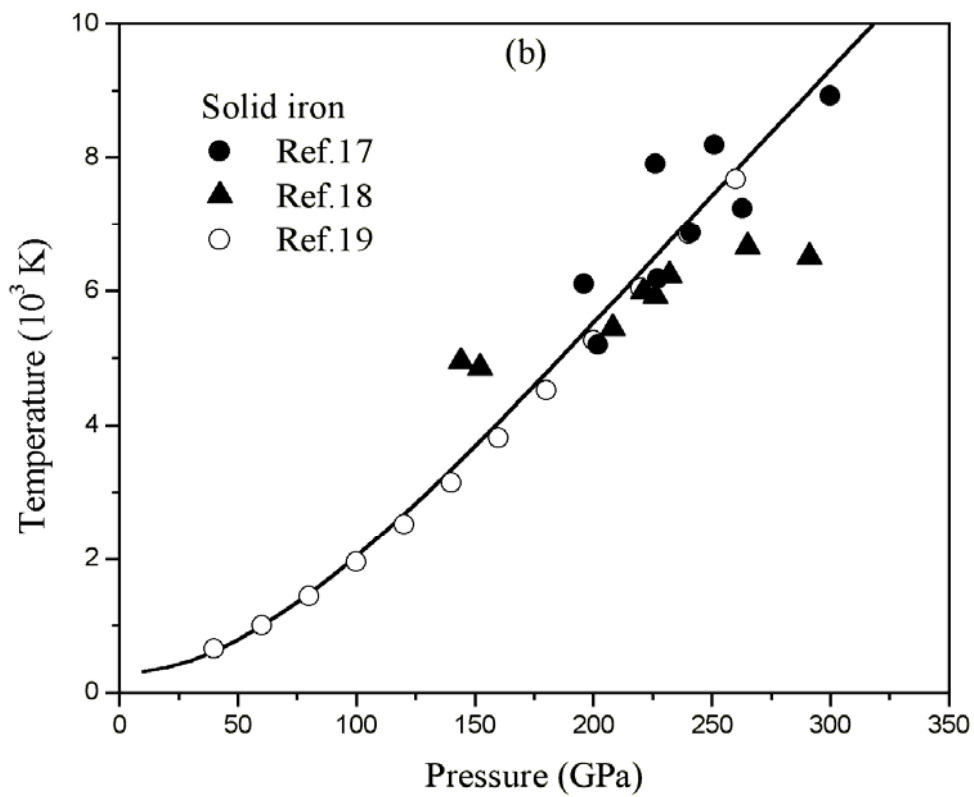

Fig.2, by Geng Huayun *et al.*

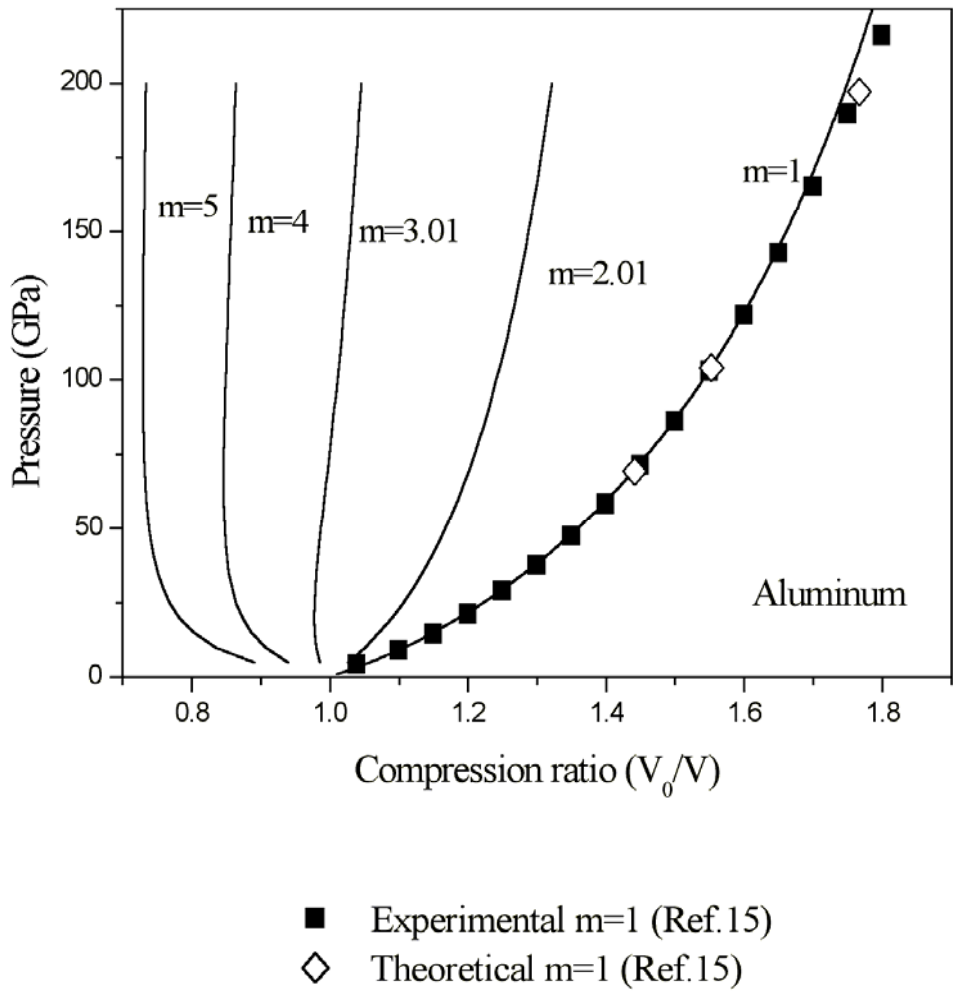

Fig.3, by Geng Huayun *et al.*

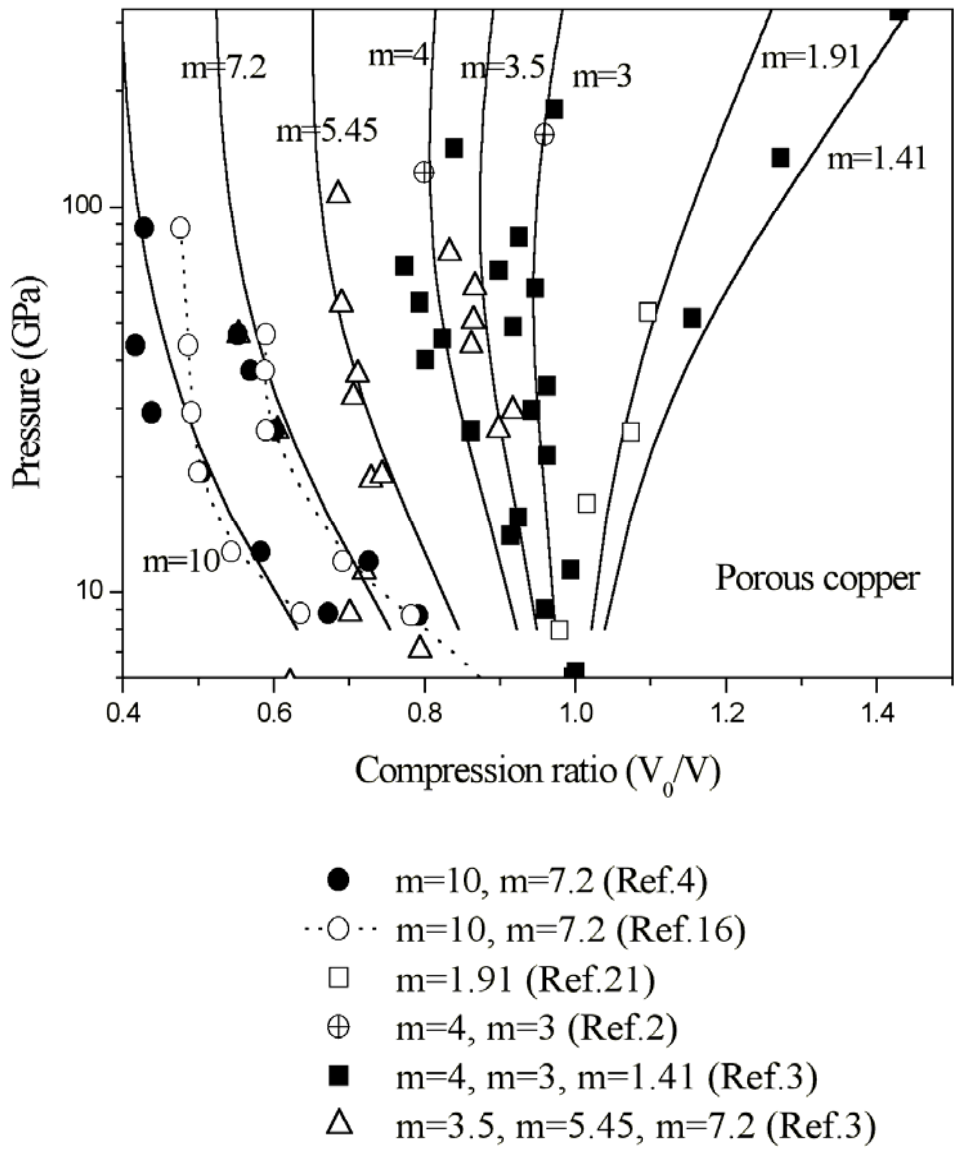

Fig.4, by Geng Huayun *et al*.

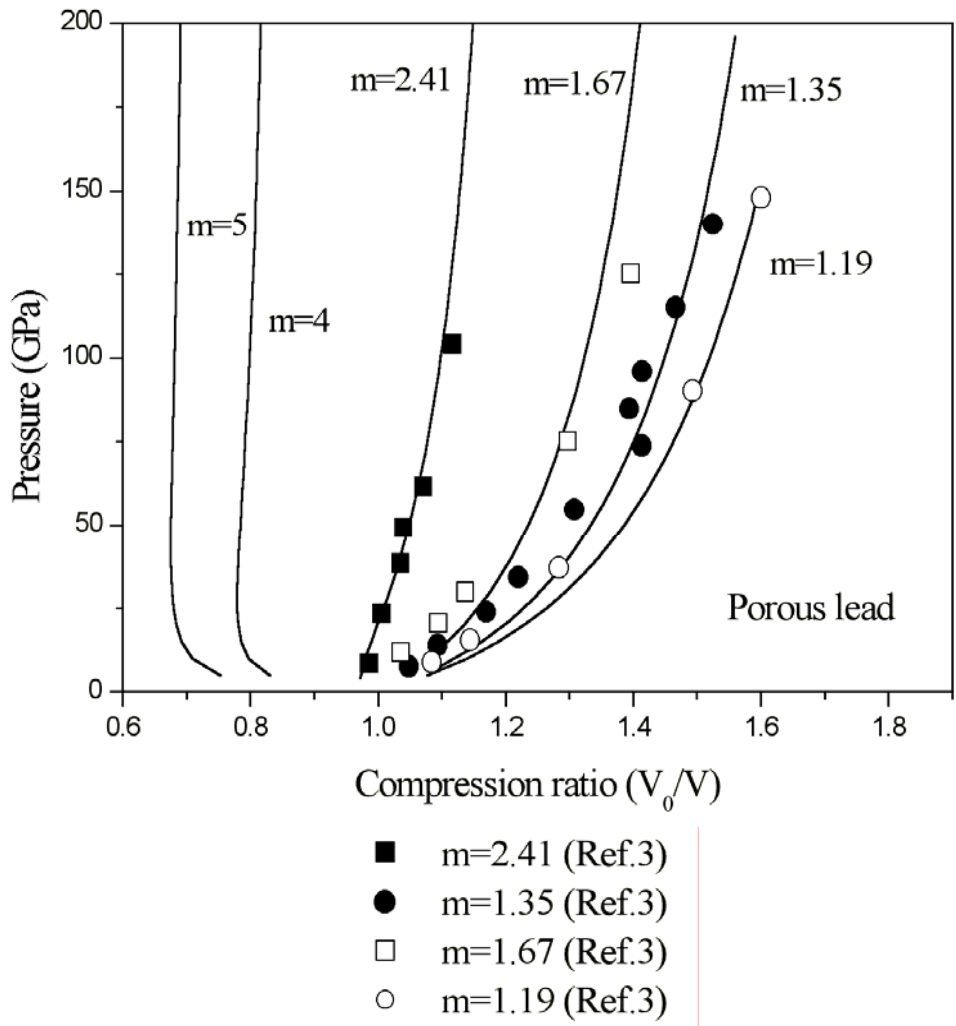

Fig.5, by Geng Huayun *et al*.

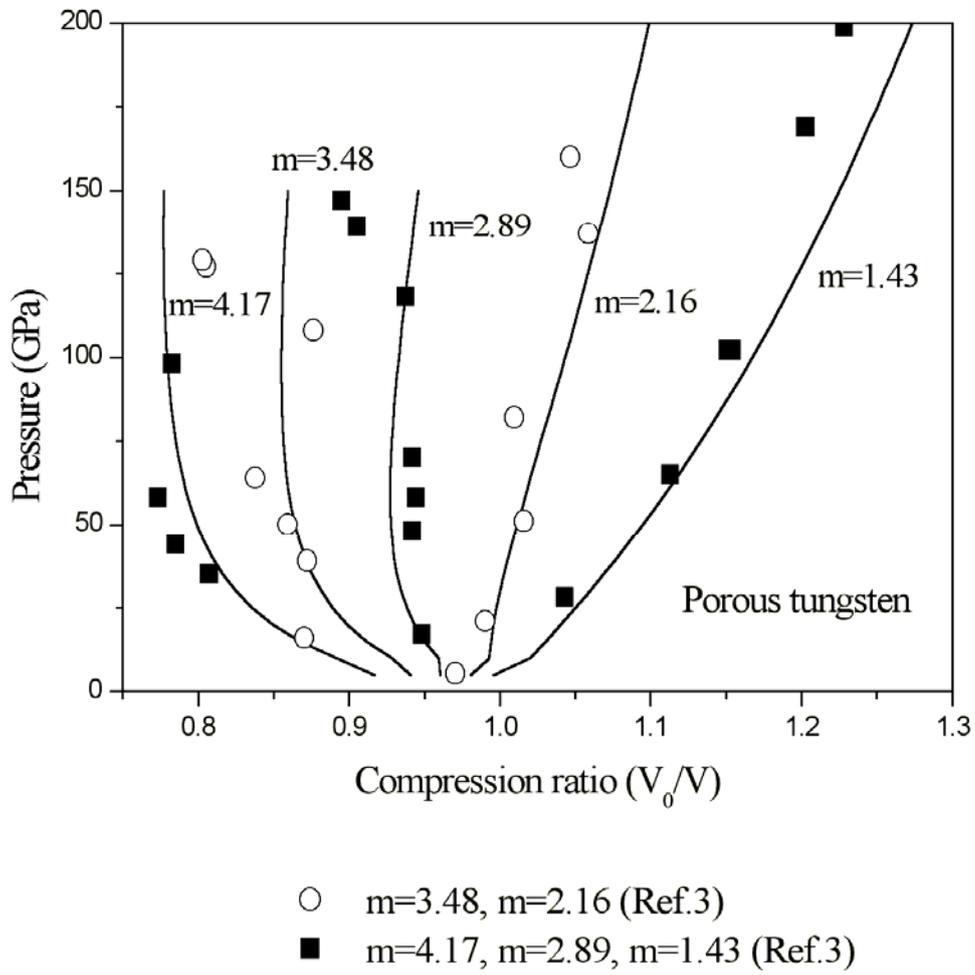

Fig.6, by Geng Huayun *et al.*

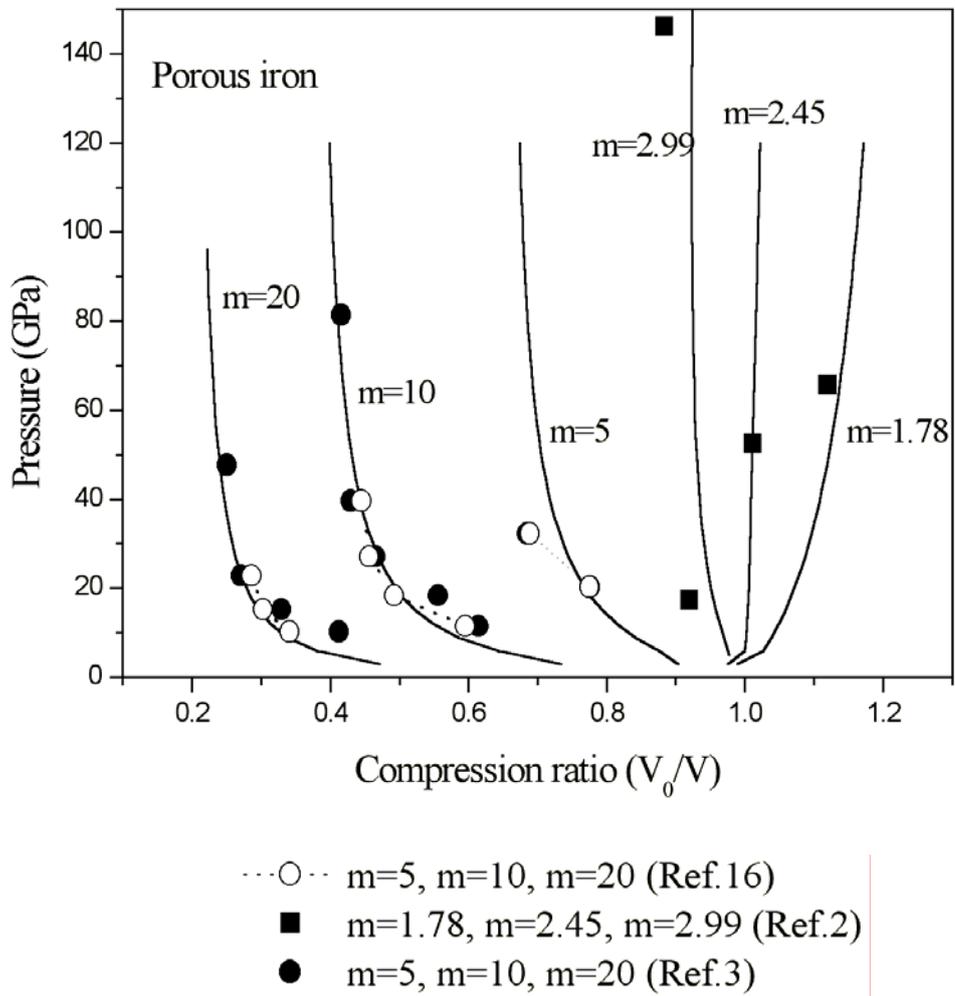

Fig.7, by Geng Huayun *et al.*

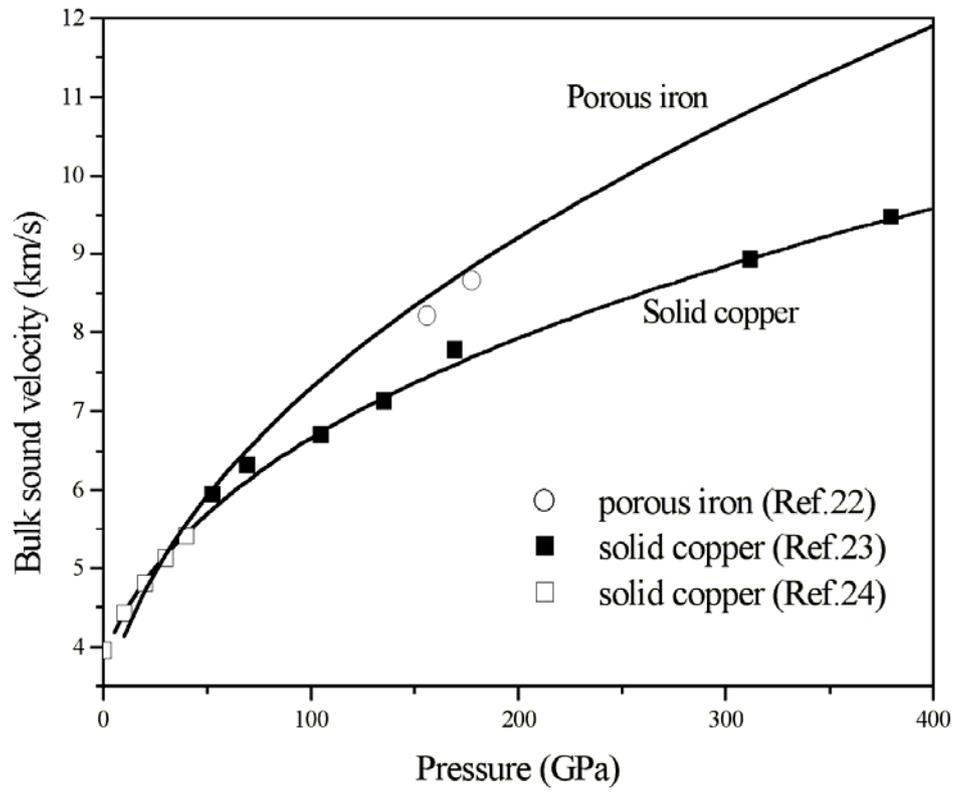

Fig.8, by Geng Huayun *et al.*

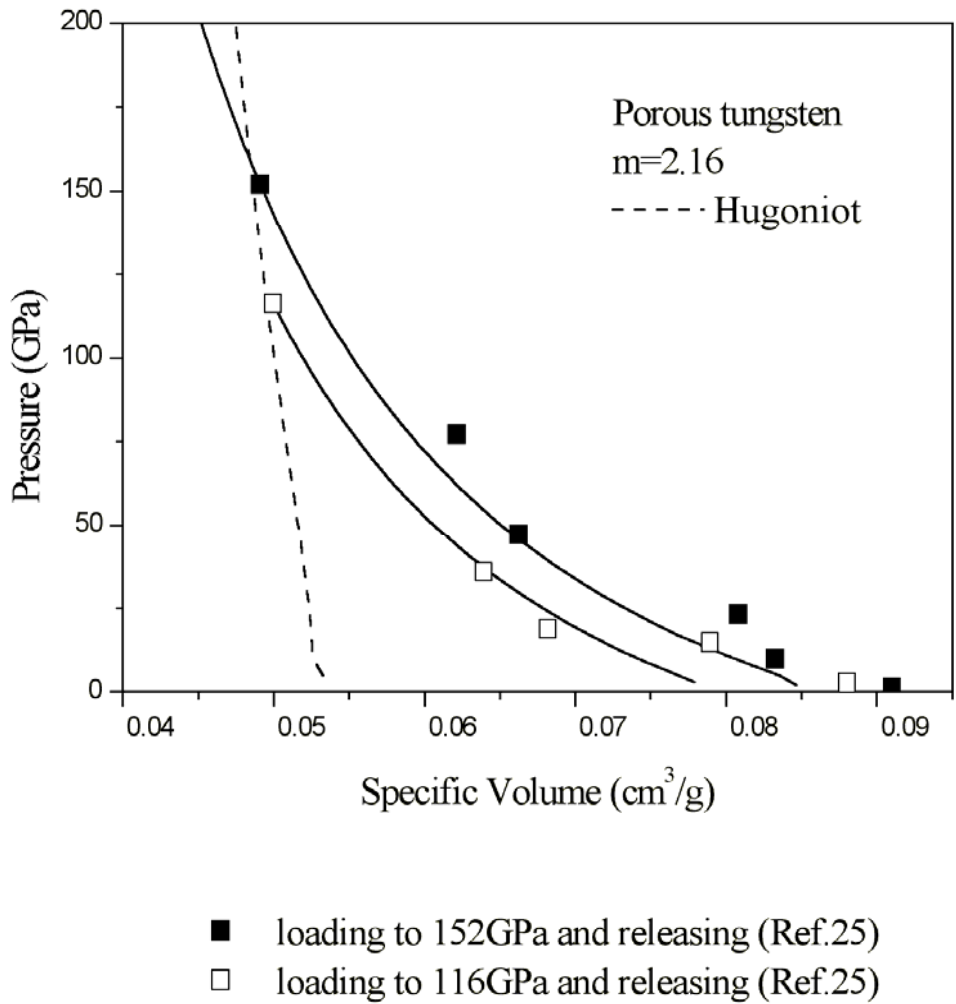

Fig.9, by Geng Huayun *et al.*

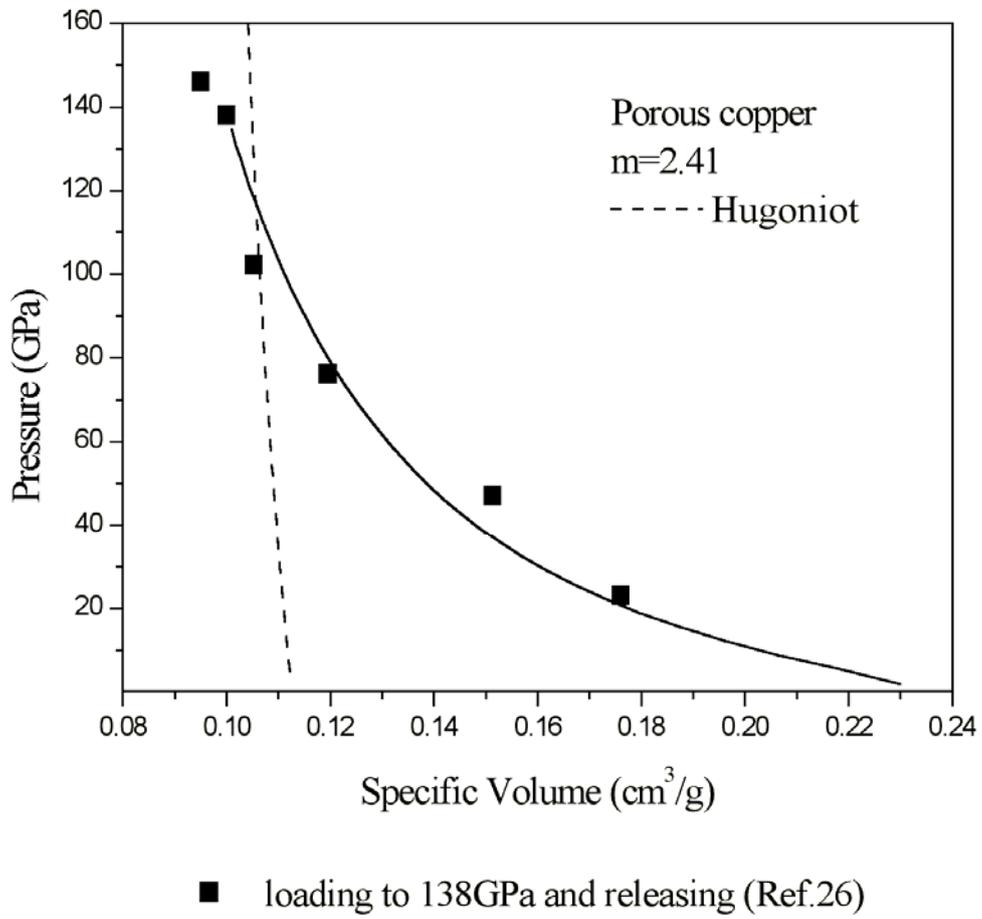

Fig.10, by Geng Huayun *et al.*

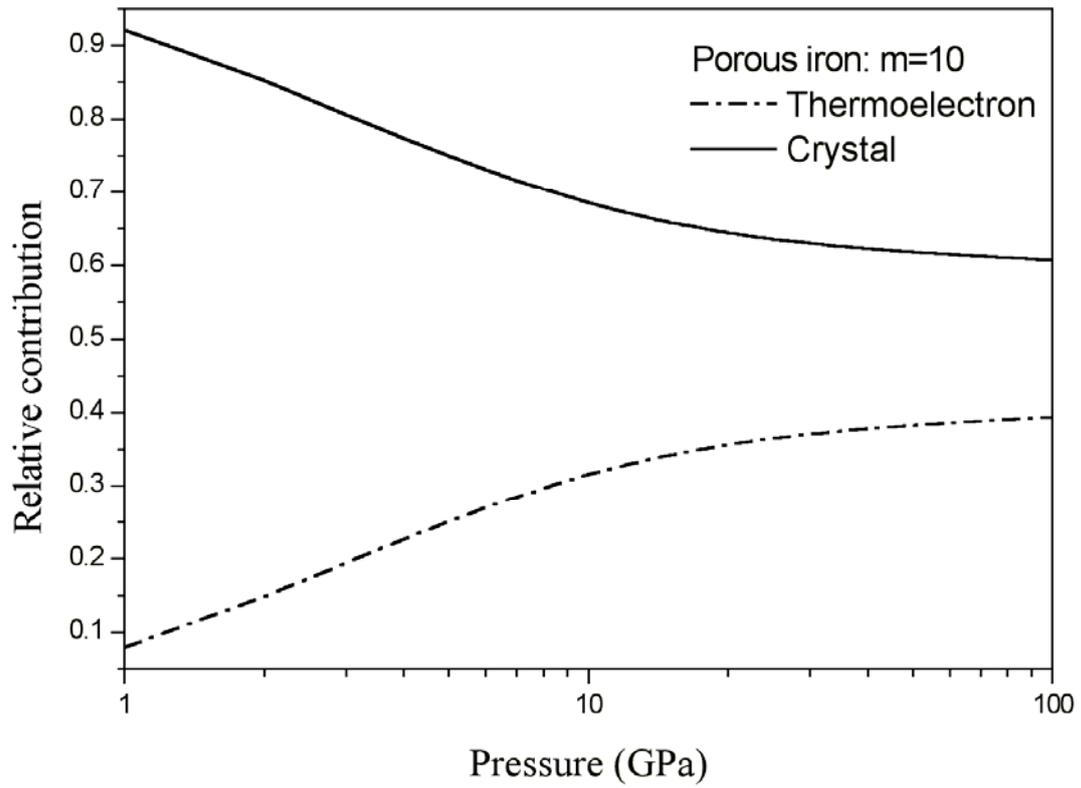

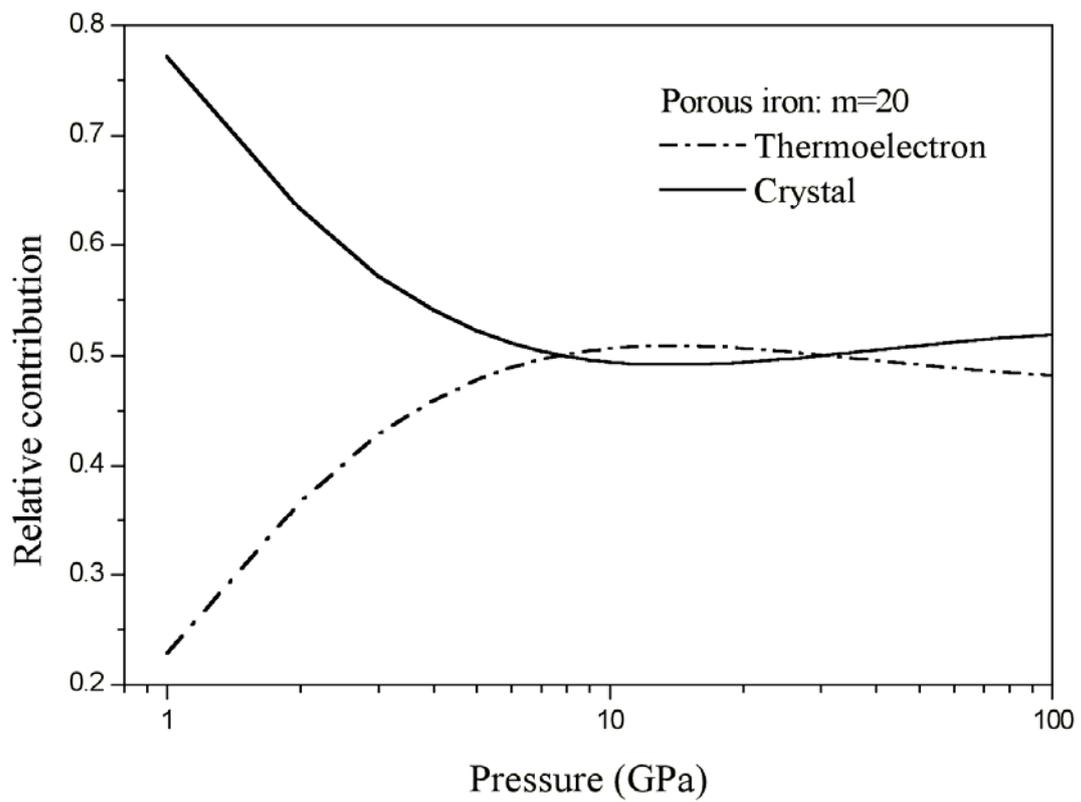

Fig.11, by Geng Huayun *et al.*